\documentclass[12pt]{article}
\usepackage{psfig}
\usepackage{graphicx}

\begin{document}

{Kinematika i Fizika Nebesnykh Tel. Vol. 22, No. 4, pp. 283-296 (2006)} \\ \\ \\ \\

\begin{center}

\bigskip

{\large \bf The Structure of the Local Supercluster of Galaxies \\ Revealed by the Three-Dimensional Voronoi's Tessellation Method
} \\

{\small O. V. Melnyk$^{1}$, A. A. Elyiv$^{1}$, I. B. Vavilova$^{2}$}  \\

{\em $^{1}$Astronomical Observatory, Kiev National University, 3 Observatorna str., \\
Kiev, 304053 Ukraine \\
melnykov@observ.univ.kiev.ua \\
$^{2}$Space Research Institute. National Space Agency of Ukraine. \\ National Academy of Sciences of Ukraine.
40 Akdemika Glushkova av., Kiev, 03680 Ukraine
}\\
\end{center}

{\large \bf Abstract}\\

3D Voronoi's tessellation method was first applied to identify groups of galaxies in the structure of a supercluster. The sample under consideration consists of more than 7000 galaxies of the Local Supercluster (LS) with radial velocities up to 3100 km s$^{-1}$. Because of an essential non-homogeneity of the LS catalogue, it was proposed to overscale distances in such an ''artificial'' way that the concentration of galaxies was varying as with increase of the distance a power-behaved function with the same exponent ${\beta}$ as for the full homogeneous catalogue. Various parameters of clustering were taking into account: ${\alpha}$ (0.01, 0.1, 1\%) as the part of galaxies, which have the relative volume of a Voronoi's cell smaller than the critical one for the random distribution; ${\beta}$ = 0, which fits to the random galaxy distribution; ${\beta}$ = 0.7, which is close to the pancake galaxy distribution. It is revealed that Voronoi's tessellation method depends weakly on ${\beta}$-parameter, and the number of galaxies in rich structures is growing rather than in poor ones with increase of ${\alpha}$-parameter. The comparison of the groups derived with the groups obtained by Karachentsev's dynamical method shows that the number of groups, which coincides by all the components, is 22\%. As a whole, the dynamical method is more preferred for identifying sparsely populated galaxy groups, whereas 3D Voronoi's tessellation method is preferred for more populated ones.


\bigskip

\section{Introduction}
The Local Supercluster (LS) of galaxies, like other superclusters, has no well-defined boundaries. The volume of the nearby Universe filled with galaxies with radial velocities approximately up to 3000 km s$^{-1}$ is meant by the LS. The LS structure has been actively studied for more than 30 years since Vaucouleurs's works [23]. Tully [19] found the LS to have three main components: 20 \% of the luminous galaxies are located in the Virgo cluster (the central part of the LS), 40 \% are concentrated in the disk, and 40 \% form a "halo"; the LS has an irregular shape. The studies by Einasto et al. [5] showed that the LS has a filamentary structure; strings of galaxies join the dominant clusters (groups) of galaxies. A series of works aimed at revealing the internal 2D and 3D structure of the LS as a whole and its individual clusters and groups appeared in the 1980s. These works, which became classical ones, primarily include the group catalog by Geller and Huchra [7] and the nearest group catalog by Tully [20], whose properties are considered in [9, 20, 21]. The tendency for luminous galaxies and low-surface brightness dwarf galaxies of the LS to clus­ter was analyzed, for example, in [4, 11].

At the same time, interest in studying and applying various galaxy clustering methods and identifying gal­axy groups has increased considerably. Thus, for example, the percolation and hierarchical galaxy group iden­tification methods suggested in [7, 20] have been the most popular clustering algorithms until now. These methods formed the basis for identifying and comparing the properties of galaxy groups in the LS volume in [6, 8, 10]. In [18], a modified percolation method was used to identify galaxy groups (v $<$ 12000 km s$^{-1}$).

A different approach to identifying LS galaxy groups was used by Makarov and Karachentsev [15]. It is based on a dynamical algorithm and allows the individual characteristics of galaxies to be completely taken into account [12]. The criteria for galaxy selection in this LS sample and the group identification procedure are described in [1, 3, 15]. Below, this catalog of LS galaxy groups will be referred to as the MK catalog [15].

The goal of this paper is to identify LS galaxy groups by an alternative method and to compare these with groups from the MK list. This will allow us to ascertain how the clustering criteria affect the charac­teristics of the groups selected from the same catalog of galaxies, on the one hand, and how the two different methods (dynamical and geometrical) agree, on the other hand. To achieve this goal, we have preferred the geometrical Voronoi tessellation method primarily because it has not yet been applied to the clustering of galaxy groups in 3D space and is more sensitive to the detection of prolate structure, which the Local Supercluster of galaxies is. Compared to this method, for example, wavelet analysis is more sensitive to the detection of spherical structures [24] and depends on the chosen size of the cluster under study, the number of galaxies in the cluster, and the distance to it [16].

The 3D Voronoi tessellation method is a geometri­cal method that uses only the positions and radial velocities of galaxies. The essence of the method is that the entire space containing galaxies is divided into ele­mentary volumes (in our case, 10 km s$^{-1}$ in size). Those of the elementary volumes that are located closer to a given galaxy than to the remaining galaxies form the volume of the Voronoi cell for this galaxy. The galaxy itself is the nucleus of this cell. The nearby galaxies with a volume smaller than the critical one form a group (Fig. 1). An automated cluster finding procedure in the 2D case is described in [17].

\begin{figure}[t]
\centerline{\includegraphics[angle=0, width=7cm]{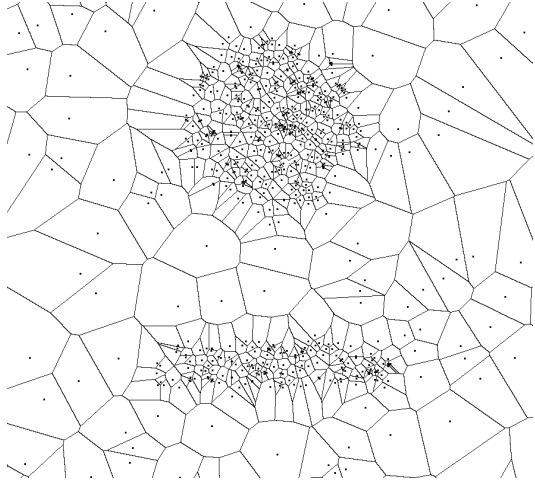}}
\caption{Example of a 2D Voronoi tessellation. 140 background points and two simulated clusters (the radii of the upper and lower clusters are 150 pcl (330 nuclei) and 480x80 pcl (150 nuclei), respectively) are randomly distributed on a surface area of 1000x1000 pcl (surfaces elements)}
\end{figure}

The Voronoi tessellation method is widely used in various fields of science, as suggested by regular inter­national meetings and conferences on the number the­ory and 3D tessellations, including those in Ukraine, where G. Voronoi was born and worked [25, 26]. For its applications to astrophysical problems, see. e.g.. [22, 25, 26].

\section{PREPARING A HOMOGENEOUS SAMPLE}

The catalog of LS galaxies under study contains 7064 galaxies with radial velocities v $<$ 3100 km s$^{-1}$ relative to the Local Group centroid [13]. Figure 2a shows the LS galaxy concentration distribution in radial velocity: the concentration decreases with depth by four orders of magnitude. Figure 2b shows the galaxy absolute magnitude distribution in radial velocity. The catalog is fairly inhomogeneous, in particular, there is no clear boundary absolute magnitude (as would be in the case of observations with a single instrument). The sample is selective in depth relative to the LS dwarf (low-luminosity) galaxies [2], The presence of a large number of dwarf galaxies in the Local Group, i.e., in the immediate neighborhood of the Galaxy, also affects the sample inhomogeneity. In addition, since the radial velocities are used as distance estimates, the supercluster is more elongated along the radial component. Calculating the Voronoi cell volumes for the gal­axies of this sample, we would obtain much smaller volumes for nearby galaxies than those for distant ones.

Therefore, it is inappropriate to directly apply the geometrical Voronoi tessellation method to such an inhomogeneous catalog. 

When the percolation and hierarchical methods are applied, this effect is taken into account by introduc­ing a luminosity function in the calculation of clustering parameters [6, 10, 18]; in this case, the sample is lim­ited in apparent magnitude $m$. In the dynamical method [12] the sample is not $m$-limited and this effect is indi­rectly taken into account in the dynamical clustering conditions (more luminous and more massive galaxies are considered to be group members, being at a larger distance from one another than less massive ones). 

Without the luminosity function and galaxy dynamics, a homogeneous catalog can be compiled by two techniques. The first is implemented by limiting the sample in absolute magnitude; $M^{0}_{abs} = -17.5^{m}$ for the LS [3]. The boundary apparent magnitude is $15.5^{m}$ and the absolute magnitude $M_{abs}$ of a galaxy with $m = 15.5^{m}$ at a distance of 3100 km s$^{-1}$ is $M^{0}_{abs}$. In this case, galaxies more luminous than $M_{abs}$ remain in the catalog (Fig. 2b).

\begin{figure}[t]
\begin{tabular}{ll}
(a) & (b) \\
\includegraphics[angle=0, width=0.48\textwidth]{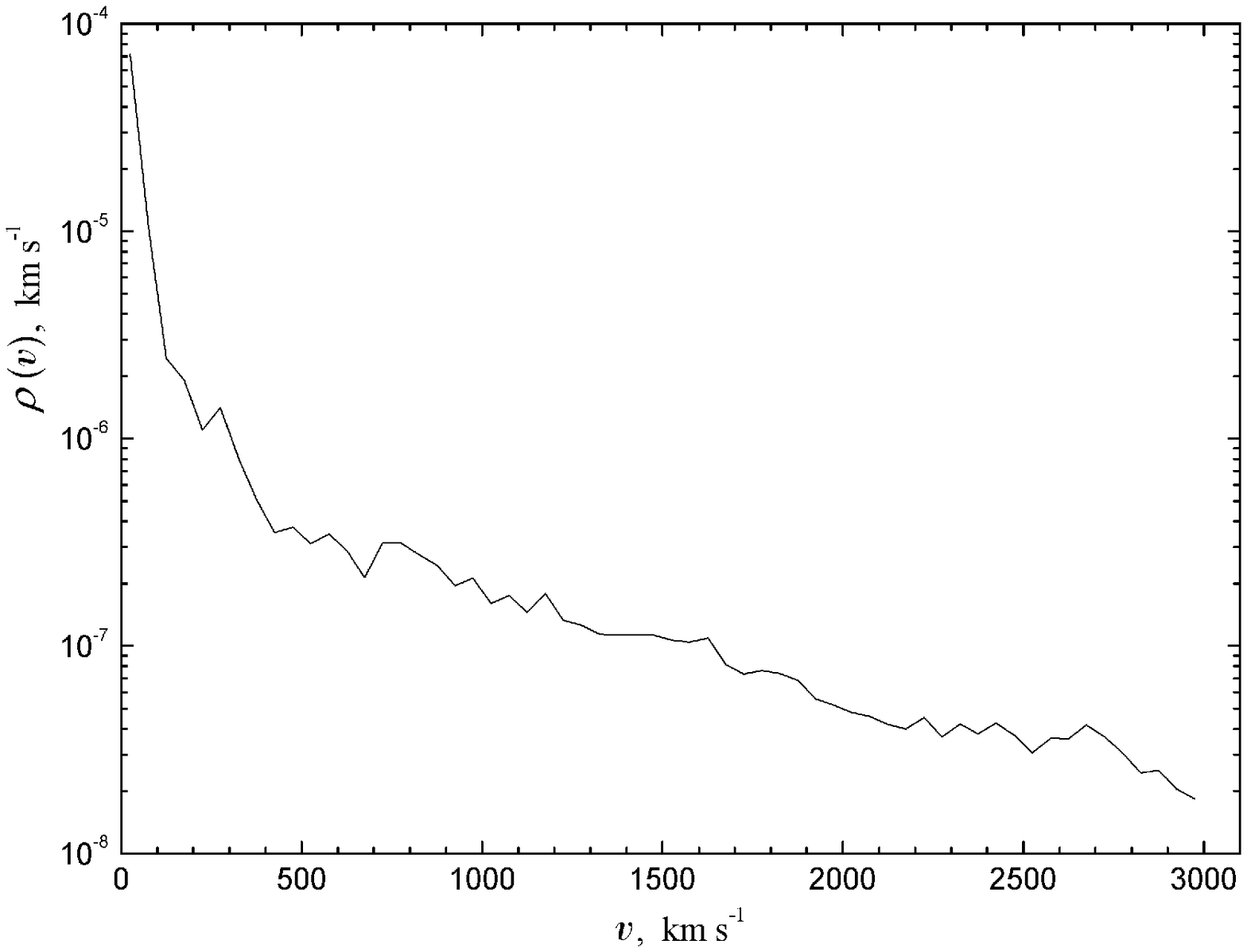} &
\includegraphics[angle=0, width=0.52\textwidth]{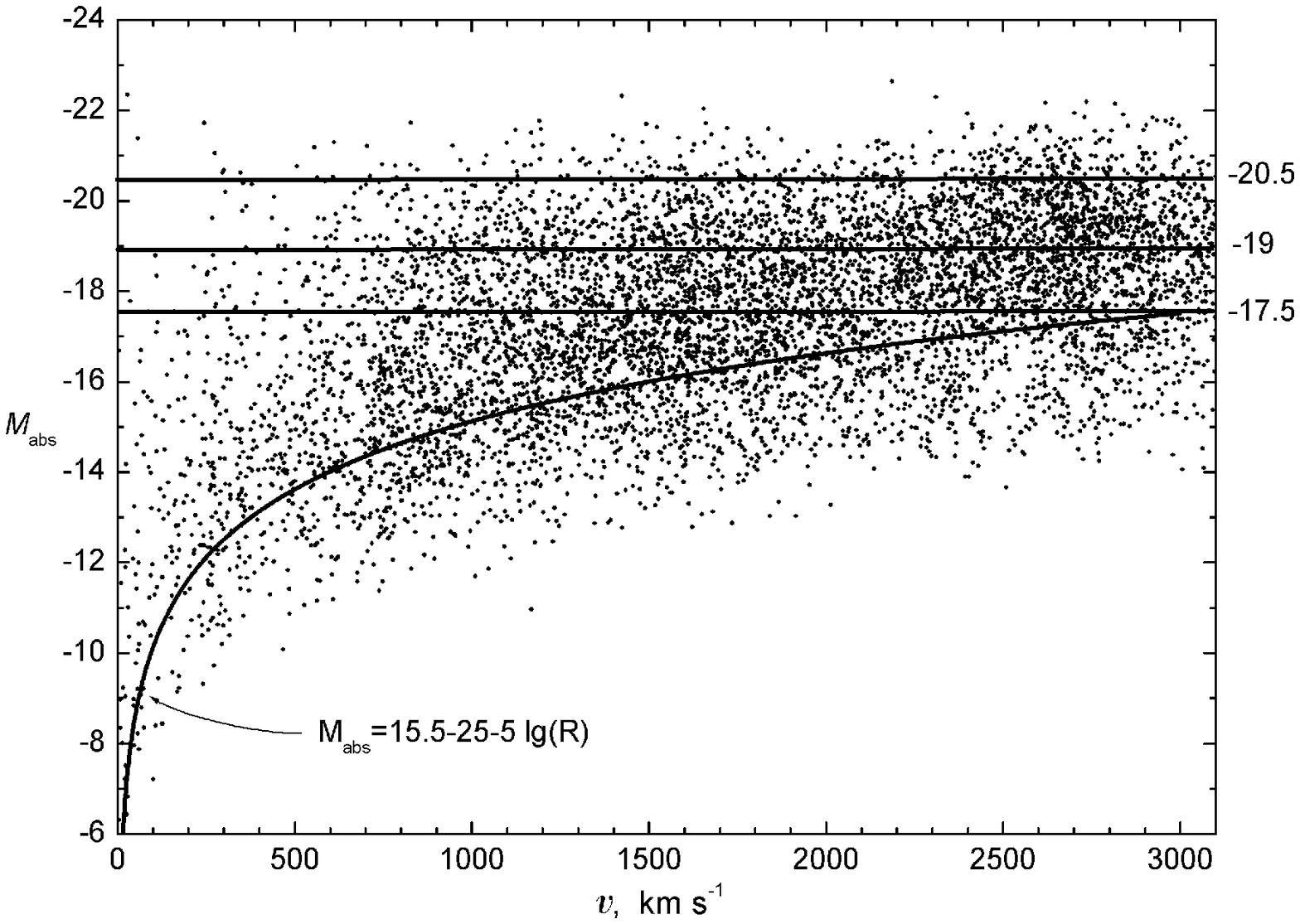} \\
\end{tabular}
\caption{LS galaxy distributions in radial velocity $v$: (a) concentration and (b) absolute magnitude. The straight lines indicate the boundary values of $M^{0}_{abs}=-17.5^{m}$, -19$^{m}$, -20.5$^{m}$}
\end{figure}

The second technique provides for the inclusion of all catalog galaxies. It involves an artificial rescaling of the distances in such a way that the concentration of galaxies varies with sample depth as a power law with the same index as that for the full homogeneous catalog. In [3], the slope was estimated for a homogeneous (in luminosity) catalog ($M_{abs} < -17.5^{m}$) to be $\beta$ = 0.7 (see below). If this value is extrapolated to the LS sam­ple under consideration, then the variation in galaxy concentration with distance will be related to the real galaxy distribution in space and not to the deficit of low-luminosity galaxies at large distances. Consider this approach in more detail.

Let us introduce a new distance as a function of the radial velocity, $u =f(v)$. The concentration of galax­ies in the ''new'' space varies as $\rho(u)=A{\cdot}u^{-\beta}$.

The concentration of galaxies at distance $v$ is 

\begin{equation}
\label{trivial}
\rho(v)=\frac{dn}{4{\pi}v^{2}dv}     
\end{equation}

in the ''old'' space and
\begin{equation}
\label{trivial}
\rho(u)=\frac{dn}{4{\pi}u^{2}du}=A{\cdot}u^{-\beta}     
\end{equation}

in the new space.

We find from Eqs. (I) and (2) that $\frac{\rho(v)}{A{\cdot}u^{-\beta}}=\frac{u^{2}}{v^{2}}\frac{du}{dv}$ or

\begin{equation}
\label{trivial}
A\int{u^{2-\beta}}du=\int{\rho}(v)v^{2}dv.
\end{equation}

The number of galaxies within the sphere of radius $v$ is

\begin{equation}
\label{trivial}
N(<v)=\int_0^v 4{\pi}v^{2}{\rho}(v)dv.
\end{equation}

Substituting (4) into (3) and integrating yields

\begin{equation}
\label{trivial}
u(v) = \left( \frac{3-{\beta}}{4{\pi}A}N(<v)\right)^{\frac{1}{3-{\beta}}}.
\end{equation}

We set the boundary condition $u(v_{max})=v_{max}$, where $v_{max}$= 3100 km s$^{-1}$ is the maximum radial velocity. Using the bound­ary condition, we obtain the constant from Eqs. (5): \\ 
$A=\frac{(3-{\beta})N}{4{\pi}v^{3-{\beta}}_{max}}$, where $N$ = 7064 is the number of galaxies in the sample. Then,

\begin{equation}
\label{trivial}
u(v)=v_{max}\left(\frac{N(<v)}{N}\right)^{\frac{1}{3-{\beta}}},
\end{equation}

This approach (distance rescaling) can be interpreted as the choice of a unit of length for the corresponding supercluster depth. It allows nearby small and distant large groups to be reduced to the same scale. Equation (6) relates the "old" dis­tance $v$ to the ''new'' distance $u$. This relation is shown in Fig. 3 for the following values of the parameter ${\beta}$: ${\beta}$ = 0, 1, and 2 cor­respond to uniform, pancake, and threadlike galaxy distribu­tions in depth, respectively. We decided to dwell on ${\beta}$ = 0.7 [3] and to consider ${\beta}$ = 0 in order to determine how sensitive the method is to this parameter.

\begin{figure}[t]
\begin{tabular}{ll}
\includegraphics[angle=0, width=0.48\textwidth]{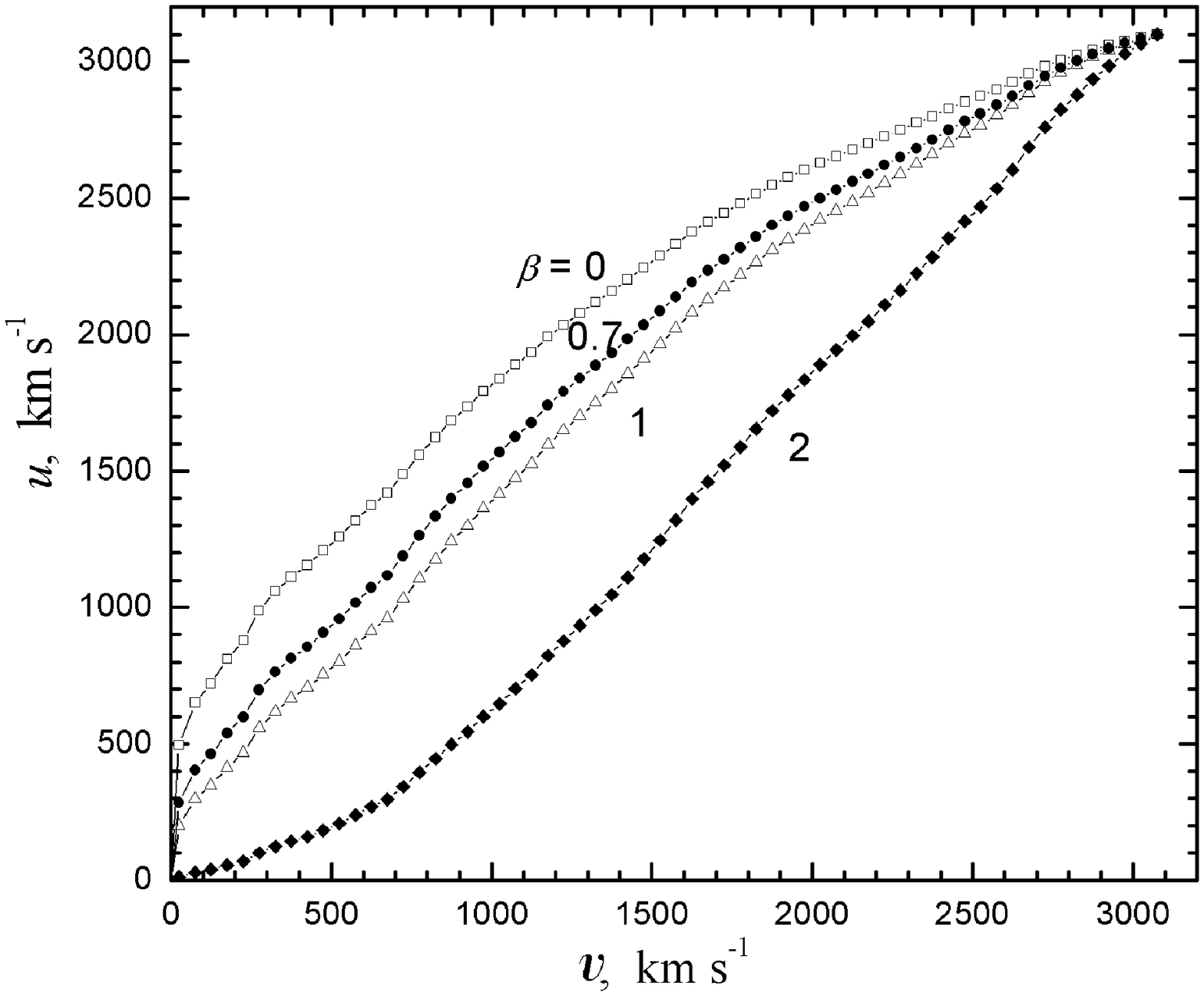} &
\includegraphics[angle=0, width=0.52\textwidth]{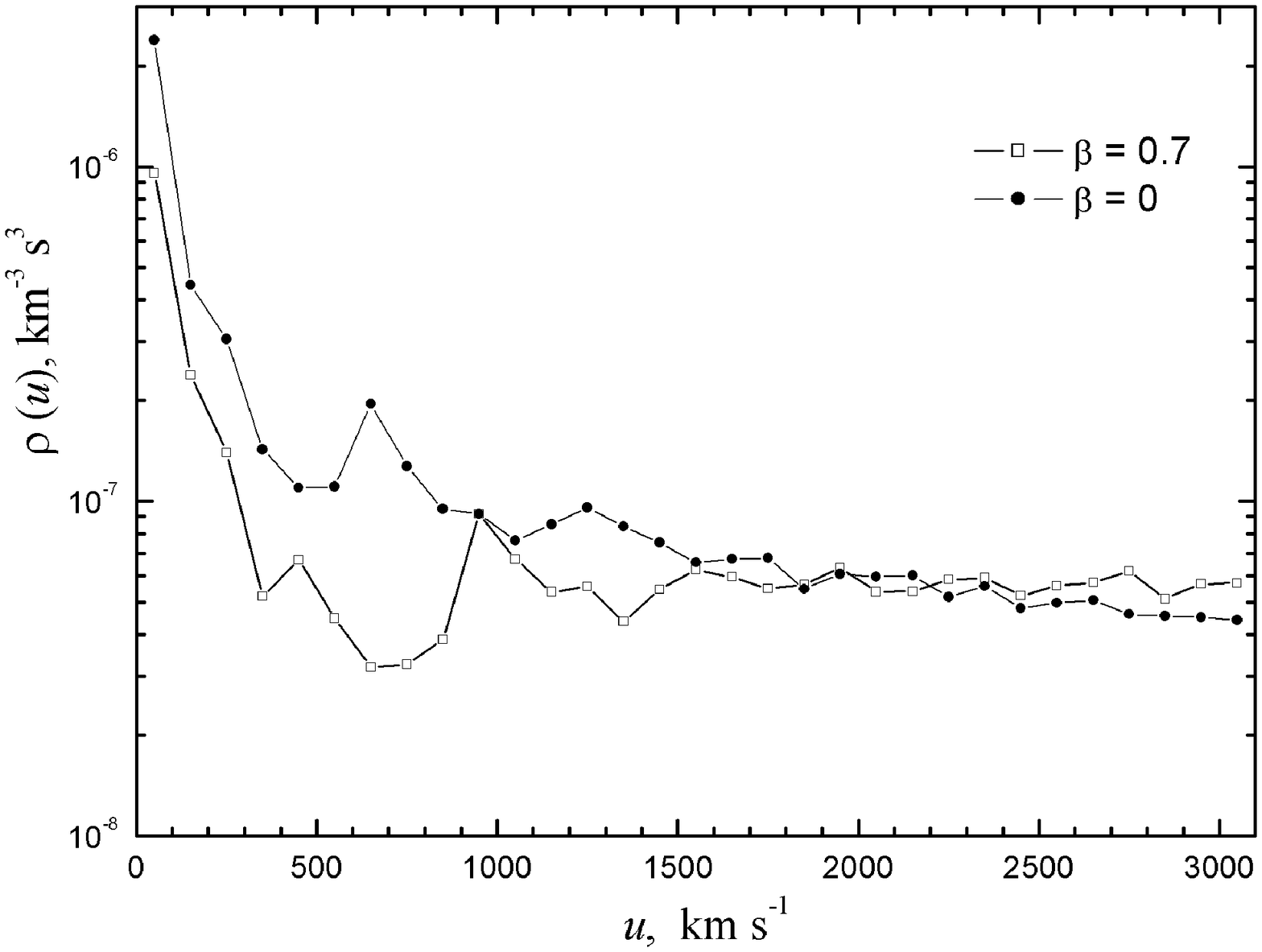} \\
\end{tabular}
\caption{New distance $u$ versus radial velocity $v$ (left).}
\caption{Galaxy concentration in the new space versus new distance $u$ (right).}
\end{figure}

In Fig. 4, the concentration of galaxies is plotted against dis­tance $u$ in the new space. The concentration now varies within one order of magnitude, in accordance with the specified slope ${\beta}$ and has no sharp negative trend toward the supercluster bound­ary. Figures 5a and 5b show the galaxy distributions in the $v$ and $u$ spaces; for comparison, Fig. 5c shows a random distribution of the same number of galaxies. Also shown are the distribu­tions of galaxies from the absolute-magnitude-limited LS subsamples: galaxies with $M_{abs} < -17.5^{m}$ (Fig. 5d), luminous galax­ies with $M_{abs} < -19^{m}$ (Fig. 5e) and the most luminous galaxies with $M_{abs} < -20.5^{m}$. Even a visual comparison demonstrates that the galaxy distributions in these subsamples are different. The distribution of the most luminous galaxies with $M_{abs} < -20.5^{m}$ is closer to the random one (but does not correspond to it). This effect in the distribution of galaxies may depend not on their luminosity, but on their number (the fewer the galaxies, the weaker their clustering (see Figs. 5 and 6)).

\begin{figure}[t]
\begin{tabular}{lll}
(a) & (b) & (c) \\
\includegraphics[angle=0, width=0.3\textwidth]{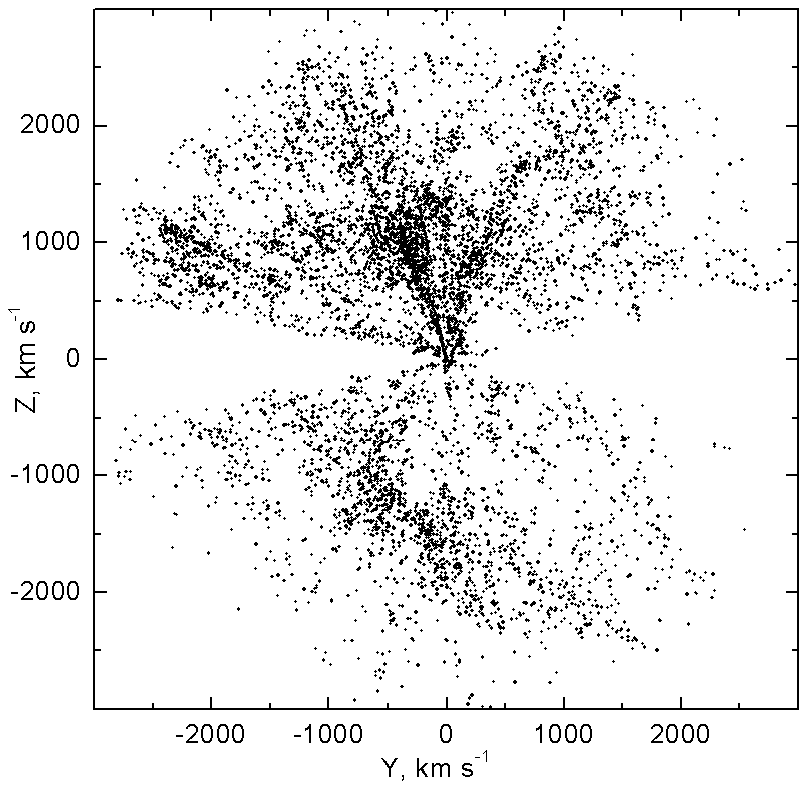} &
\includegraphics[angle=0, width=0.3\textwidth]{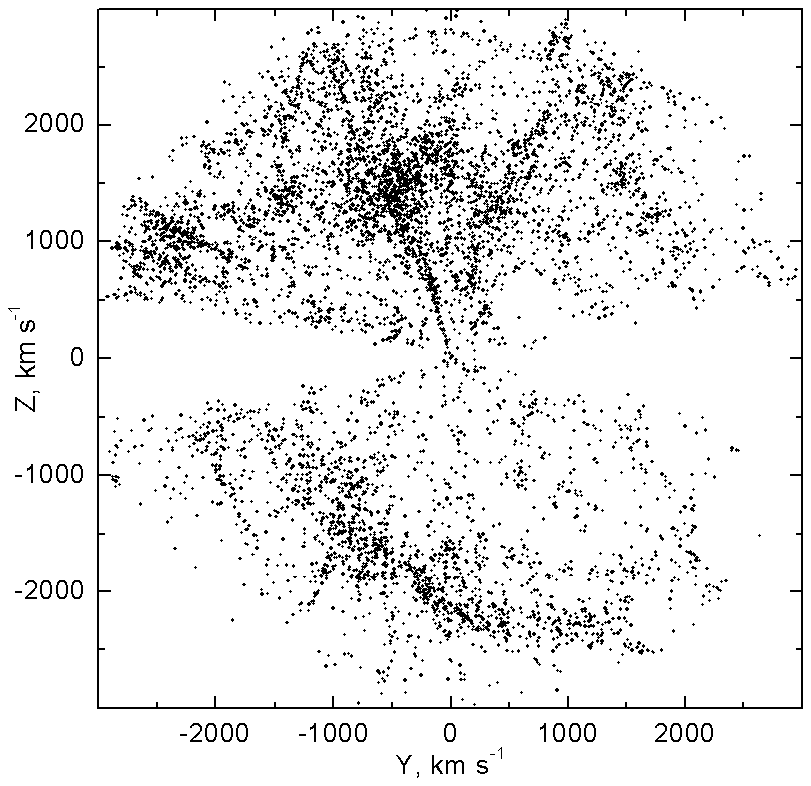} &
\includegraphics[angle=0, width=0.3\textwidth]{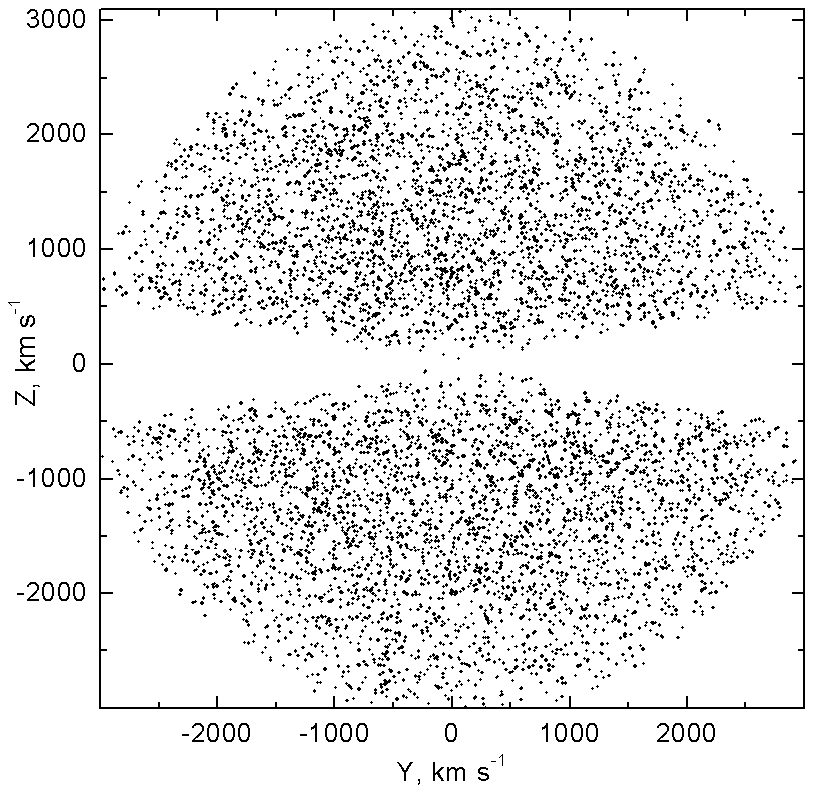} \\
(d) & (e) & (f)\\
\includegraphics[angle=0, width=0.3\textwidth]{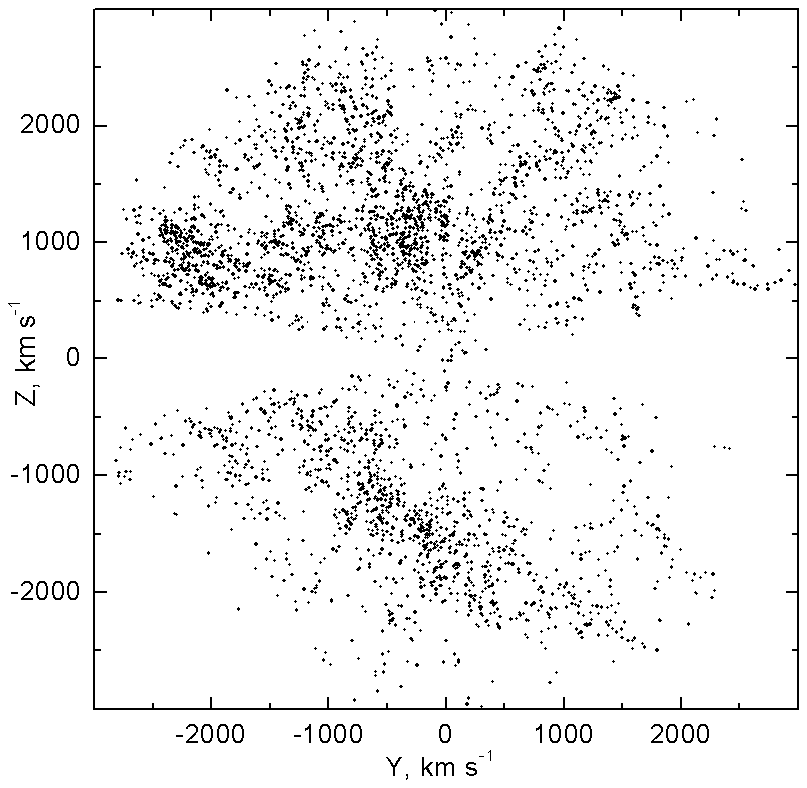} &
\includegraphics[angle=0, width=0.3\textwidth]{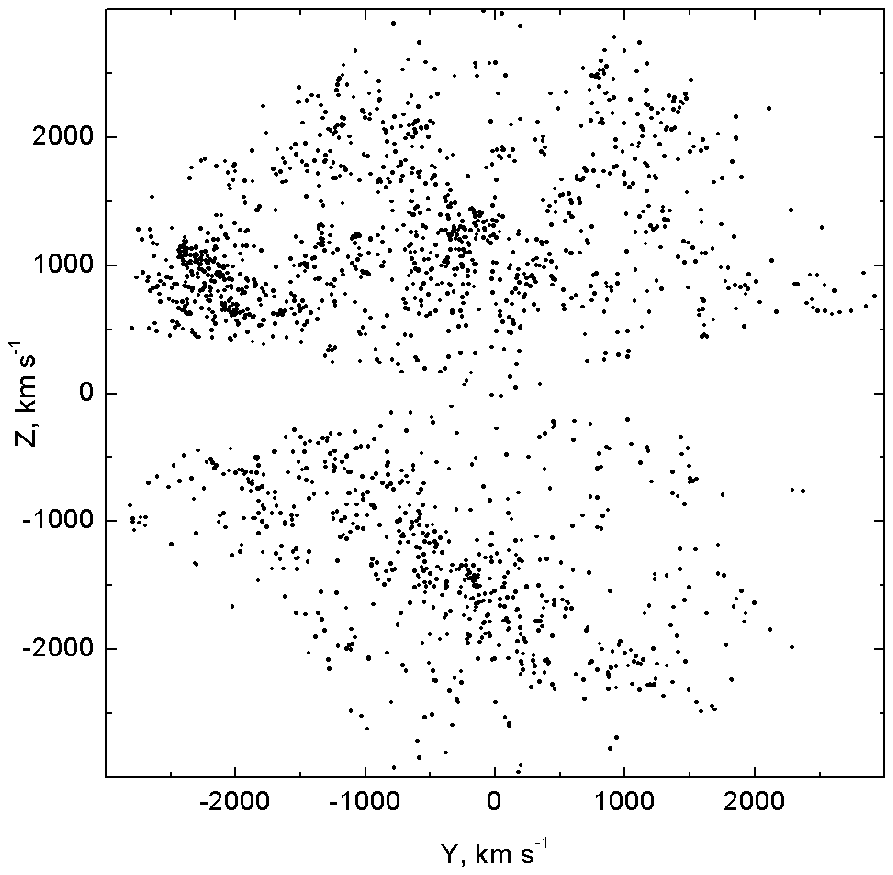} &
\includegraphics[angle=0, width=0.3\textwidth]{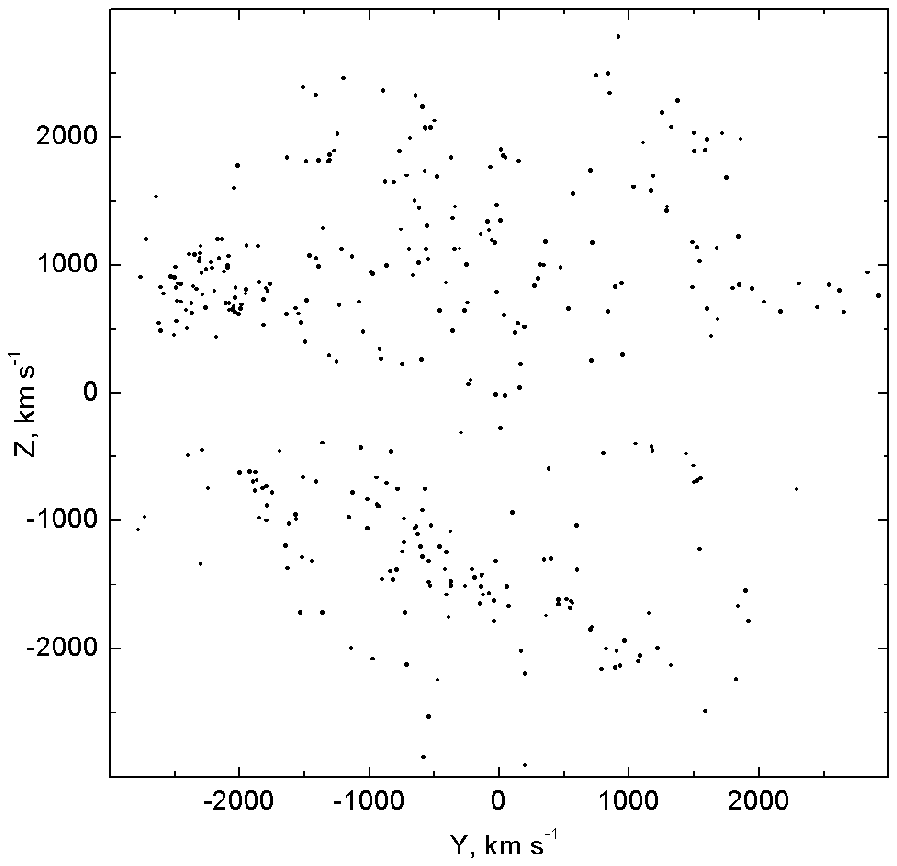} \\
\end{tabular}
\caption{Projection of the LS sample galaxies onto the plane perpendicular to the galactic plane: (a) in the $v$ space, (b) in the $u$ space, (c) a random distribution of the same number of galaxies, (d) the distribution of LS galaxies with $M_{abs} < -17.5^{m}$ in the $v$ space, (e) the distribution of LS galaxies with $M_{abs} < -19^{m}$ in the $v$ space, (f) the distribution of LS galaxies with $M_{abs} < -20.5^{m}$ in the $v$ space.}
\end{figure}

Let us analyze these distributions in more detail and identify LS galaxy groups by the 3D Voronoi tes­sellation method.

\section{SEARCING FOR GALAXY GROUPS BY  \\ THE VORONOI TESSELLATION METHOD}

According to Kiang [14], the density of the Voronoi cell volume distribution for a random distribution of nuclei is analytically described by the function

\begin{equation}
\label{trivial}
H(x,c)=\frac{c}{\Gamma(c)}(cx)^{c-1}e^{-cx},                                                            
\end{equation}

where $x = V/{\langle}V{\rangle}$, $V$ is the Voronoi cell volume, ${\langle}V{\rangle}$ is the mean Voronoi cell volume, and $\Gamma(c)$ is the Gamma function; $c$ = 6 for the 3D case. The galaxy group identification procedure consists in the following: the galaxies for which the Voronoi cell volume is smaller than a certain boundary relative volume $x_{lim}$ will be considered to be clustered ones. As the boundary value, we will take such $x_{lim}$, that the fraction ${\alpha}$ of galaxies for a random distribution has a Voronoi cell volume smaller than $x_{lim}$. Figure 6 shows the cumulative Voronoi cell volume distribution func­tions for a random distribution according to (7) and for the LS subsamples under consideration. For ${\alpha}$ = 0.1 \%, $x_{lim}$  = 0.1825. The remaining galaxies will acquire the status of ''nonclustered'' ones for the boundary Voronoi cell volume in question. There are galaxies among them that are group members, but their cells have a volume that exceeds $x_{lim}$ since these are located at the group boundary (see Fig. 1. where boundary cells of large size are clearly seen for both clusters). Only the boundary galaxies the distances from which to the nearest clustered galaxy were smaller than the characteristic radius of the largest cell, $R = 0.5d\sqrt[3]{x_{lim}{\langle}V{\rangle}}$ , where $d$ = 10 km s$^{-1}$ is the chosen size of the elementary LS volume, were added to the groups. There are also ''superfluous'' galaxies with a relative Voronoi cell volume smaller than the critical one in the sample of clustered galaxies, but there is no galaxy at a distance smaller than $R$ near them. Following the criterion described above, we added the ''boundary'' galaxies to the sample of clustered galaxies and discarded the superfluous ones. Thus, using the 3D Voronoi tessellation method, we divided the sample into two parts: galaxies in groups and nonclustered galaxies.

\begin{figure}[t]
\begin{tabular}{ll}
(a) & (b) \\
\includegraphics[angle=0, width=0.49\textwidth]{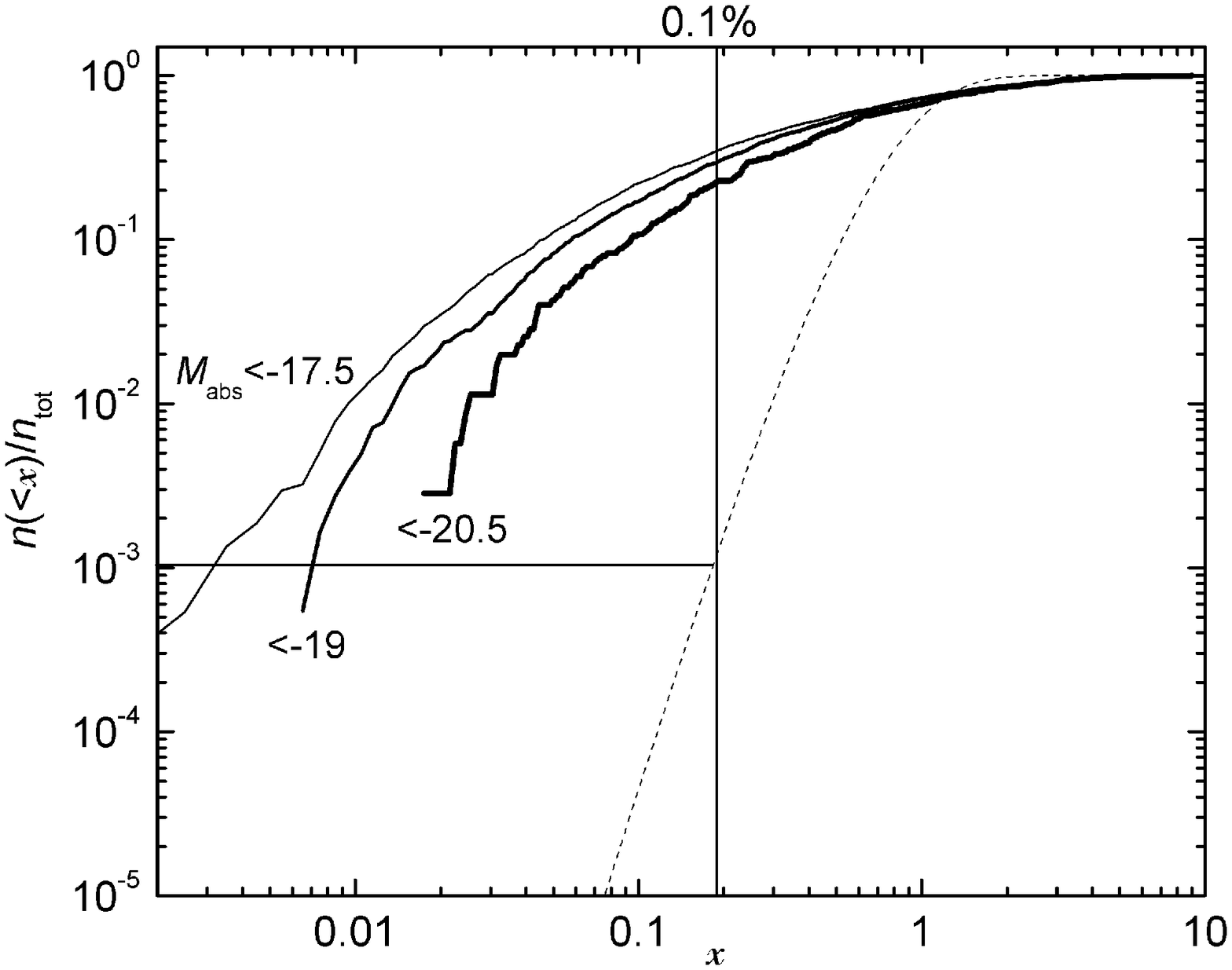} &
\includegraphics[angle=0, width=0.5\textwidth]{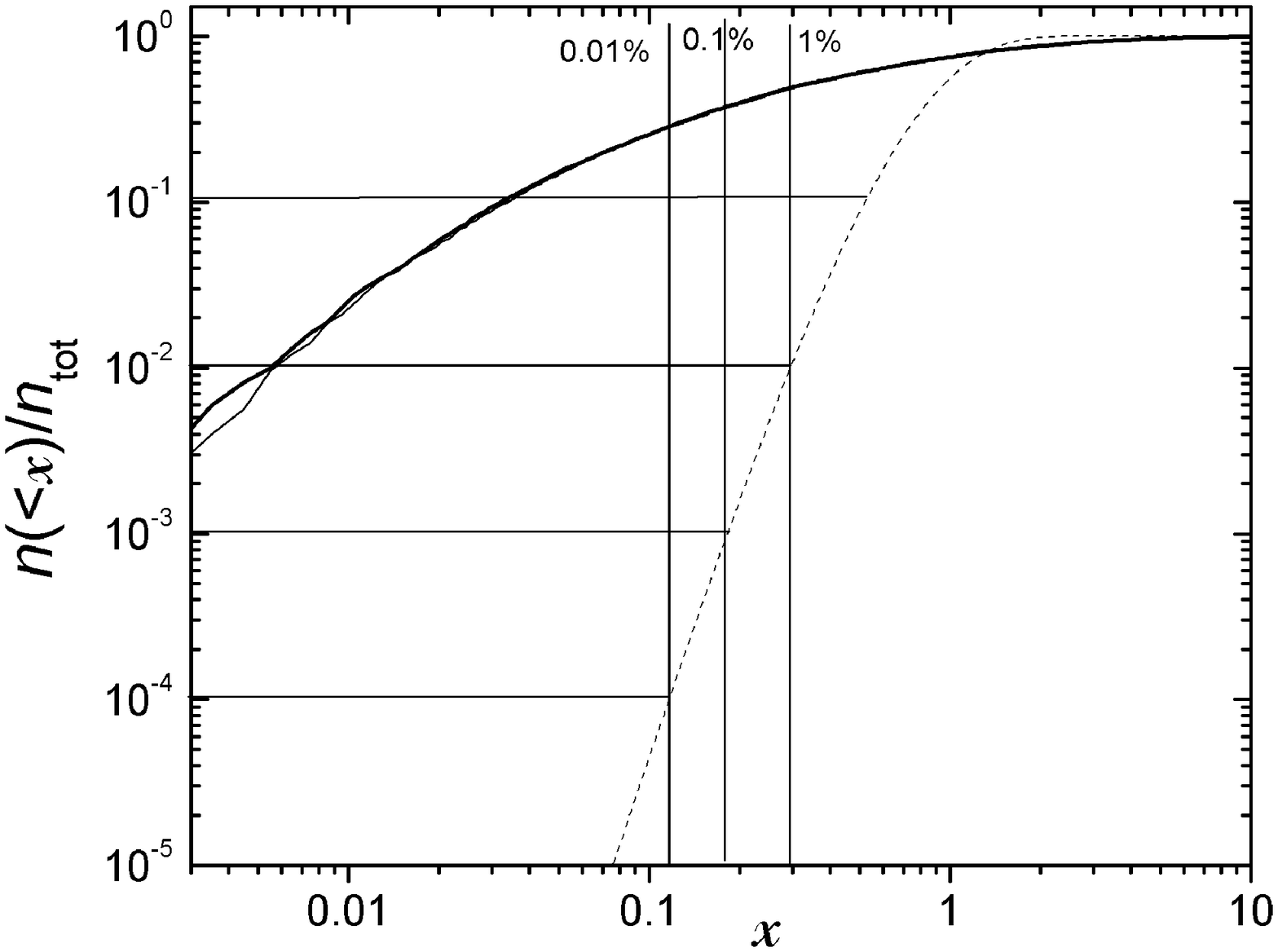} \\
\end{tabular}
\caption{Comparison of the cumulative observed distributions of the relative volumes $x$ with a random distribution of relative Voronoi cell volumes (dotted line): (a) for $M_{abs}$-limited subsamples and (b) for subsamples corresponding to different ${\beta}$.}
\end{figure}

Below, we present the results of applying the 3D Voronoi tessellation method to the homogeneous sam­ples obtained by the techniques considered in the previous section:

(1) Let us first consider three absolute-magnitude-limited LS samples (Fig. 2b): $M_{abs} < -17.5^{m}$, -19$^{m}$, and -20.5$^{m}$ (in these cases, no distance rescaling is needed to achieve catalog homogeneity). Table 1 presents the following parameters: $N$ is the number of galaxies in the sample, ${\langle}V{\rangle}$ is the mean Voronoi cell volume (in arbitrary units), $R$ is the characteristic radius of the largest cell. $N_{gr}$ is the number of galaxies in groups, and $N_{ncl}$ is the number of nonclustered galaxies. The cumulative distributions of relative Voronoi cell volumes $x$ for these three samples are compared with a random distribution in Fig. 6a: at small $x$, the real galaxy distribution differs sharply from the random one; the more luminous the galaxies remain in the $M^{0}_{abs}$ -limited subsample, the closer the volume distribution of their cells to the random one.

(2) Let us apply the 3D Voronoi tessellation method to the full sample of LS galaxies rescaled according to (6), where ${\beta}$ = 0 and 0.7 and ${\langle}V{\rangle}$ = 14597. Normalizing the volumes of all cells to this value, we obtain the cumulative distribution (Fig. 6b). Clearly, the real distribution at small $x$ differs greatly from the random one, as in case 1. Consider three values of $x_{lim}$ at which 0.01, 0.1, and 1 \% of the galaxies for a random dis­tribution will have a relative volume $x$ smaller than $x_{lim}$. The clustering parameters corresponding to ${\alpha}$ = 0.01, 0.1, and 1 \% - are presented in Table 2. Note that the char­acteristic radius of the largest cell $R$ is a constant in the $u$ space; in the real $v$ space, this parameter varies with radial velocity over the range from 10 to 350 km s$^{-1}$ (Fig. 7).

\parskip=3 mm
Table 1. Clustering parameters for the $M^{0}_{abs}$-limited subsamples of LS galaxies.

\begin{tabular}{|l|l|l|l|l|l|}
\hline
$M^{0}_{abs}$ & $N$ & ${\langle}V{\rangle}$ & $R$, km s$^{-1}$ & $N_{gr}$ & $N_{ncl}$\\
\hline
-17.5& 3730 & 27645 & 85.8 & 1637 (43.9 \%) & 2093 (56.1 \%)\\
\hline
-19 &1826 & 56471 & 108.8 & 718 (39.3 \%) & 1108 (60.7 \%)\\
\hline
-20.5 &351 &293778 & 188.5 & 112 (31.9 \%) & 239 (68.1 \%)\\
\hline
\end{tabular}

\parskip=3 mm
Table 2. Clustering parameters for the full sample of the LS galaxies at various ${\alpha}$ and ${\beta}$.

\begin{tabular}{|l|l|l|l|l|l|l|}
\hline
${\alpha}$, \% & $x_{lim}$ & $R$, km s$^{-1}$ & $N_{gr}$  (${\beta} = 0$) & $N_{ncl}$ (${\beta} = 0$) & $N_{gr}$  (${\beta} = 0.7$) & $N_{ncl}$ (${\beta} = 0.7$)\\
\hline
0.01& 0.115 & 59.4 & 2523 (35.7 \%) & 4541 (64.3 \%)& 2497 (35.3 \%) & 4567 (64.7 \%)\\
\hline
0.1 & 0.1825 & 69.3 & 3306 (46.8 \%) & 3758 (53.2 \%)& 3309 (46.8 \%) & 3755 (53.2 \%)\\
\hline
1 &0.297 & 81.5 & 4094 (58.0 \%) & 2970 (42.0 \%)& 4126 (58.4 \%) & 2938 (41.6 \%)\\
\hline
\end{tabular}

\parskip=3 mm

\begin{figure}[t]
\centerline{\includegraphics[angle=0, width=9cm]{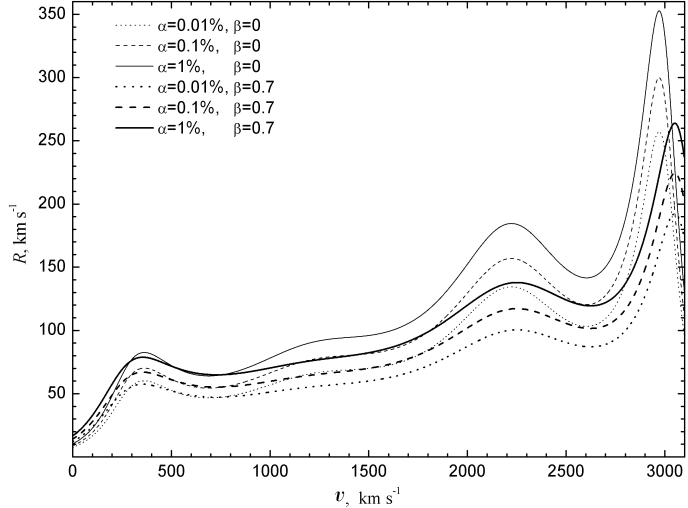}}
\caption{Variation in the characteristic radius of the Voronoi cell corresponding to the volume $x_{lim}$ in the $v$ space.}
\end{figure}

Since the set of galaxies with a distance between them smaller than the clustering radius $R$ is meant by the group of galaxies, the groups identified by the first technique will have, on average, larger sizes than those identified by the second technique (Tables 1 and 2). As follows from Table 1, the smaller the number of galaxies is represented in the $M_{abs}$ -limited subsamples, the larger the fraction of non-clustered galaxies ($N_{ncl}$). In other words, the more lumi­nous the galaxies remain in the subsample, the weaker their tendency to cluster. As a result, the groups revealed by the first technique cannot be compared with the groups identified by the second technique.

Since, as we noted in the Introduction, our goal is also to compare the groups selected from the same catalog of LS galaxies by alternative methods, below only the groups identified by the second techniques at different parameters ${\alpha}$ and ${\beta}$ will be compared with the MK groups.

\section{IDENTIFYING THE DERIVED GROUPS WITH MK GROUPS}

We obtained six samples of groups selected by the geometrical Voronoi tessellation method (below these samples will be referred to as MEV groups) at various parameters ${\alpha}$ and ${\beta}$. Let us compare them with the MK sample of groups selected by the dynamical method [12]. Table 3 shows the number of MEV galaxy groups $n^{tot}_{gr}$ at given ${\alpha}$ and ${\beta}$, the number of MK groups, and the fraction of coincided groups $n_{gr}/n^{tot}_{gr}$.

\parskip=3 mm

Table 3. Comparison of the numbers of MEV and MK groups.

\begin{tabular}{|l|l|l|l|l|l|l|l|}
\hline
MEV & \multicolumn{3}{c|}{${\beta} = 0$} & \multicolumn{3}{c|}{${\beta} = 0.7$} & MK\\
\hline
${\alpha}$, \% & 0.01 & 0.1 & 1 & 0.01 & 0.1 & 1 & -\\
\hline
$n_{gr}^{tot}$ & 557 & 608 & 644 & 531 & 605 & 645 & 929 \\
\hline
$n_{gr}/n_{gr}^{tot}$ &0.21 & 0.22 & 0.23 & 0.18 & 0.22 & 0.22 & -\\
\hline
\end{tabular}

\parskip=3 mm

\begin{figure}[t]
\begin{tabular}{ll}
(a) & (b) \\
\includegraphics[angle=0, width=0.535\textwidth]{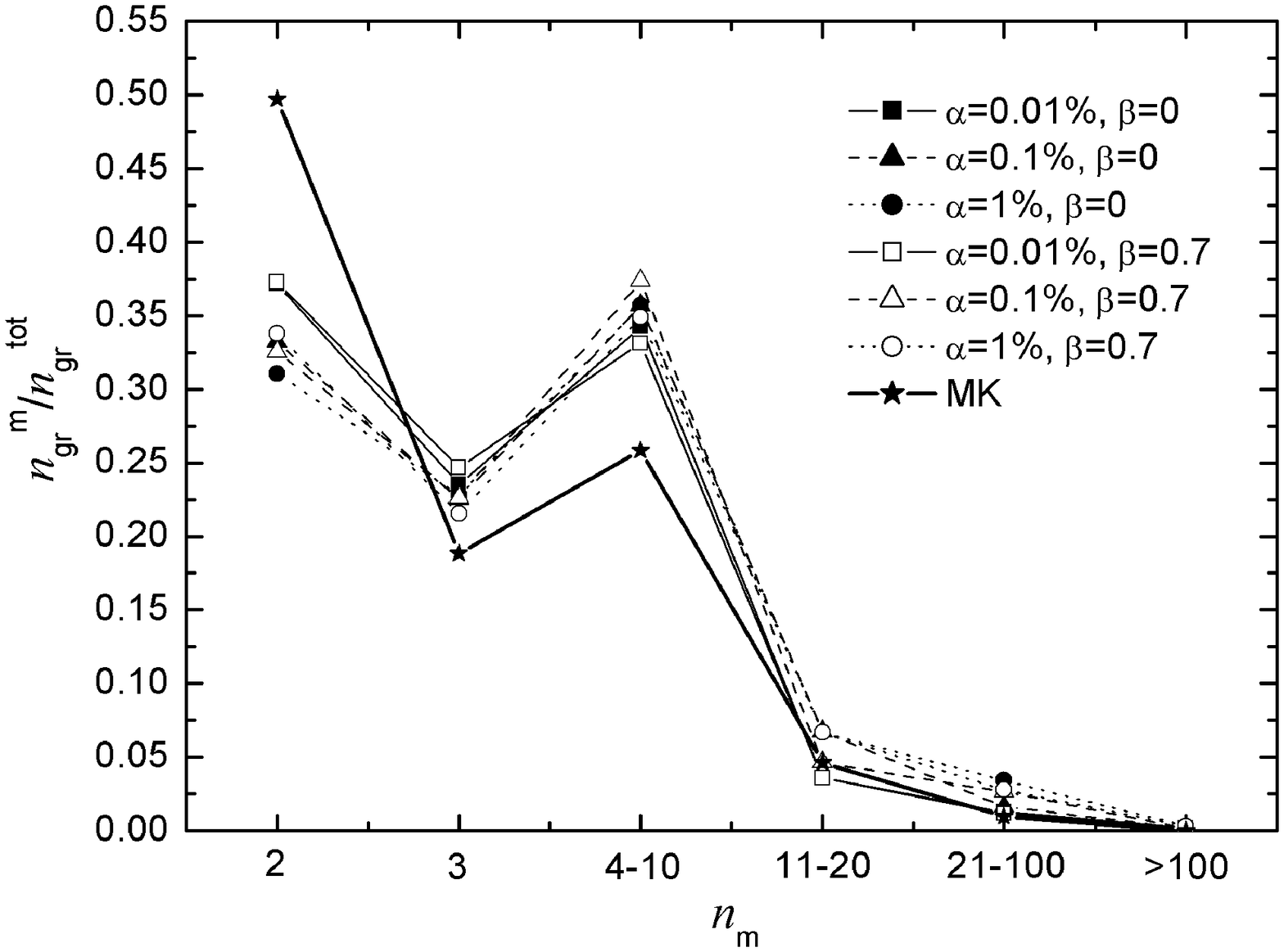} &
\includegraphics[angle=0, width=0.465\textwidth]{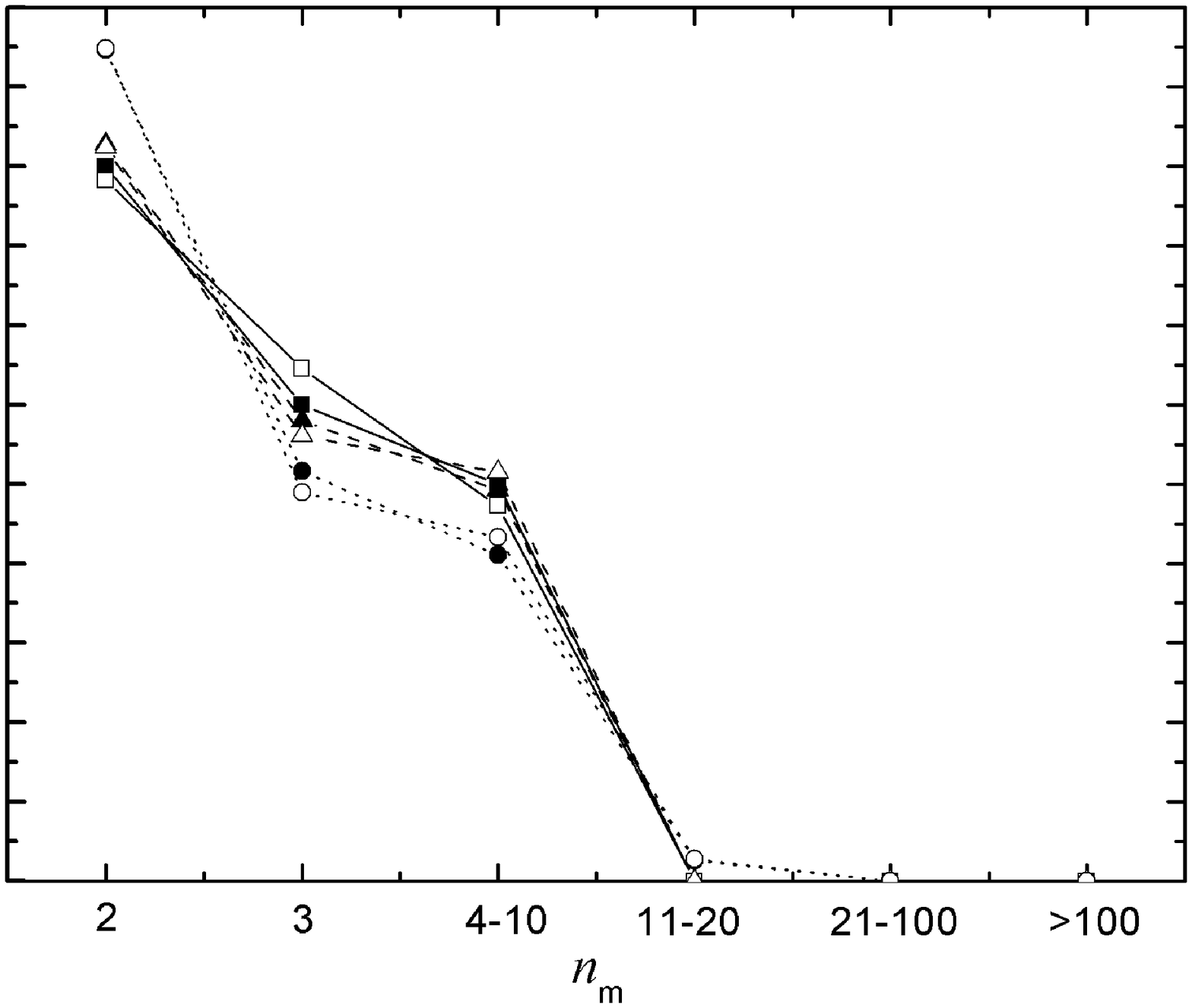} \\
\end{tabular}
\caption{Distributions of (a) the relative number of galaxy groups in component number in the group for various parameters ${\alpha}$ and ${\beta}$ and (b) the relative number of MEV groups coincided with the MK groups.}
\end{figure}

The largest relative number of MEV groups coincide with the MK groups at ${\alpha}$ = 0.1 and 1 \%: the fraction of coincided groups is 22 \% irrespective of ${\beta}$. Figure 8a presents the group number distributions correspond­ing to different numbers of components $n_{m}$ in the group for all samples; $n^{m}_{gr}/n^{tot}_{gr}$, is the ratio of the number of groups at given $n_{m}$ to the total number of groups in the sample under consideration. Figure 8b presents the distributions of coincided groups al given $n_{m}$.

\begin{figure}[t]
\begin{tabular}{ll}
(a) & (b) \\
\includegraphics[angle=0, width=0.535\textwidth]{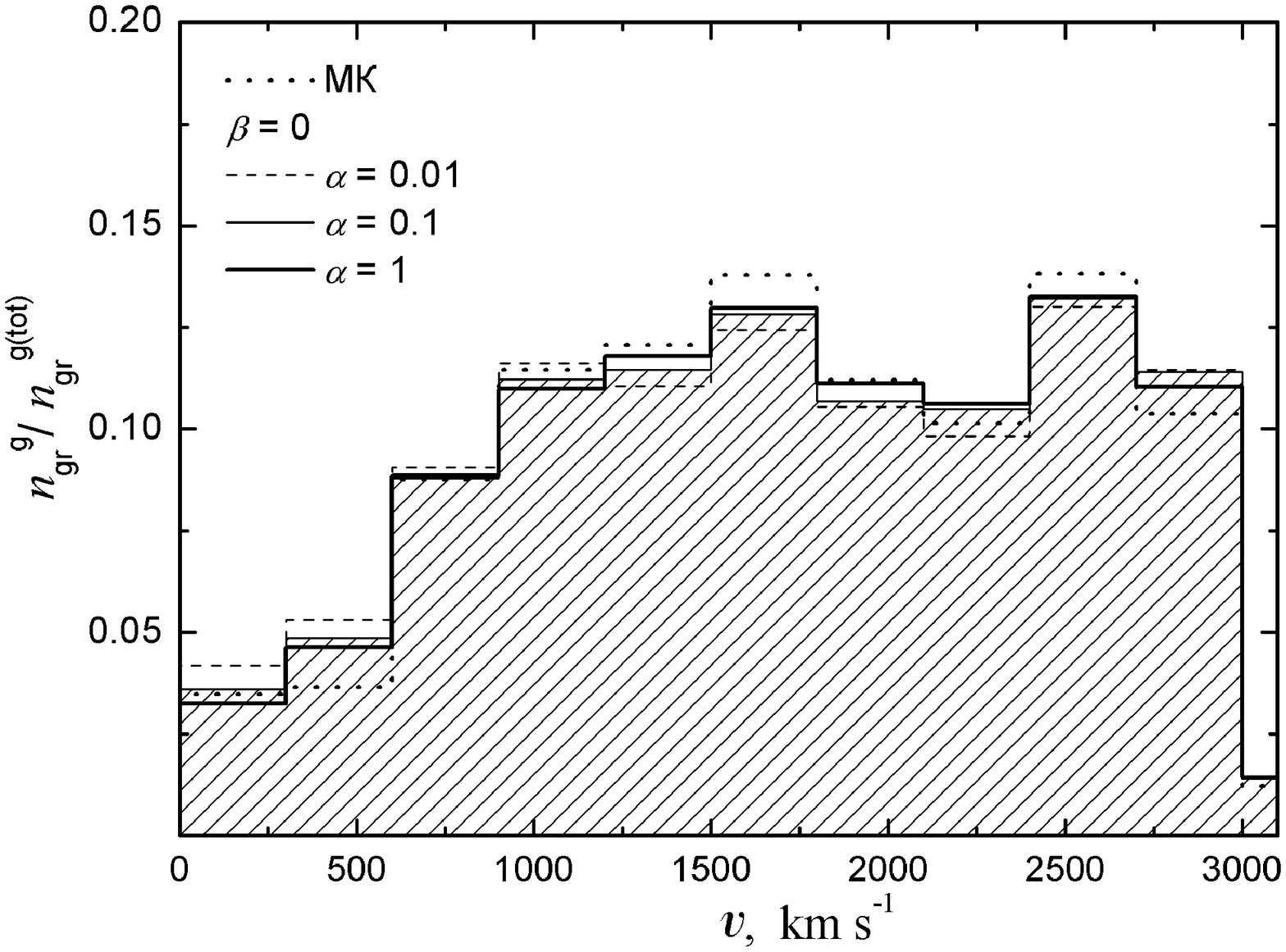} &
\includegraphics[angle=0, width=0.465\textwidth]{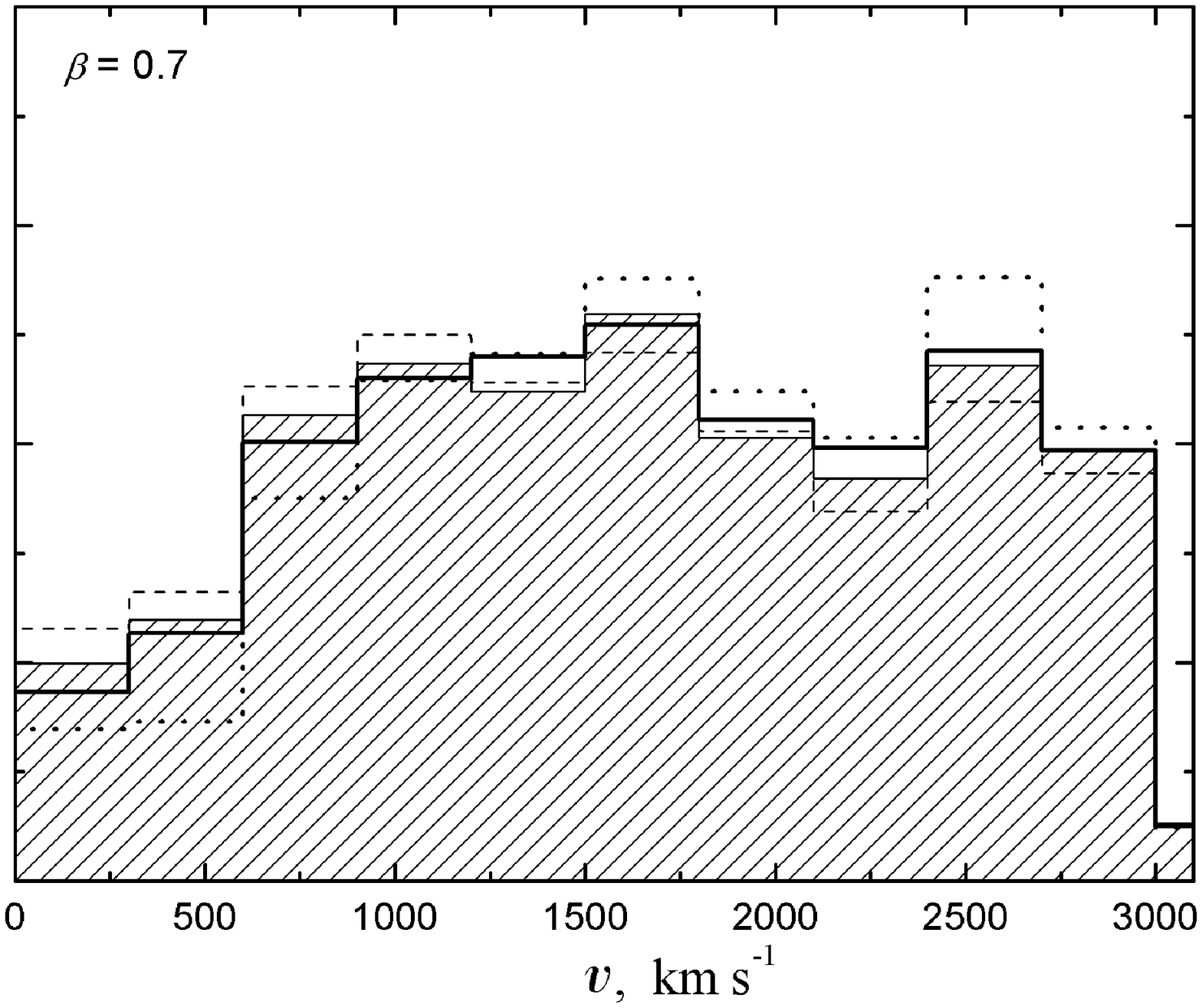} \\
(c) & (d) \\
\includegraphics[angle=0, width=0.535\textwidth]{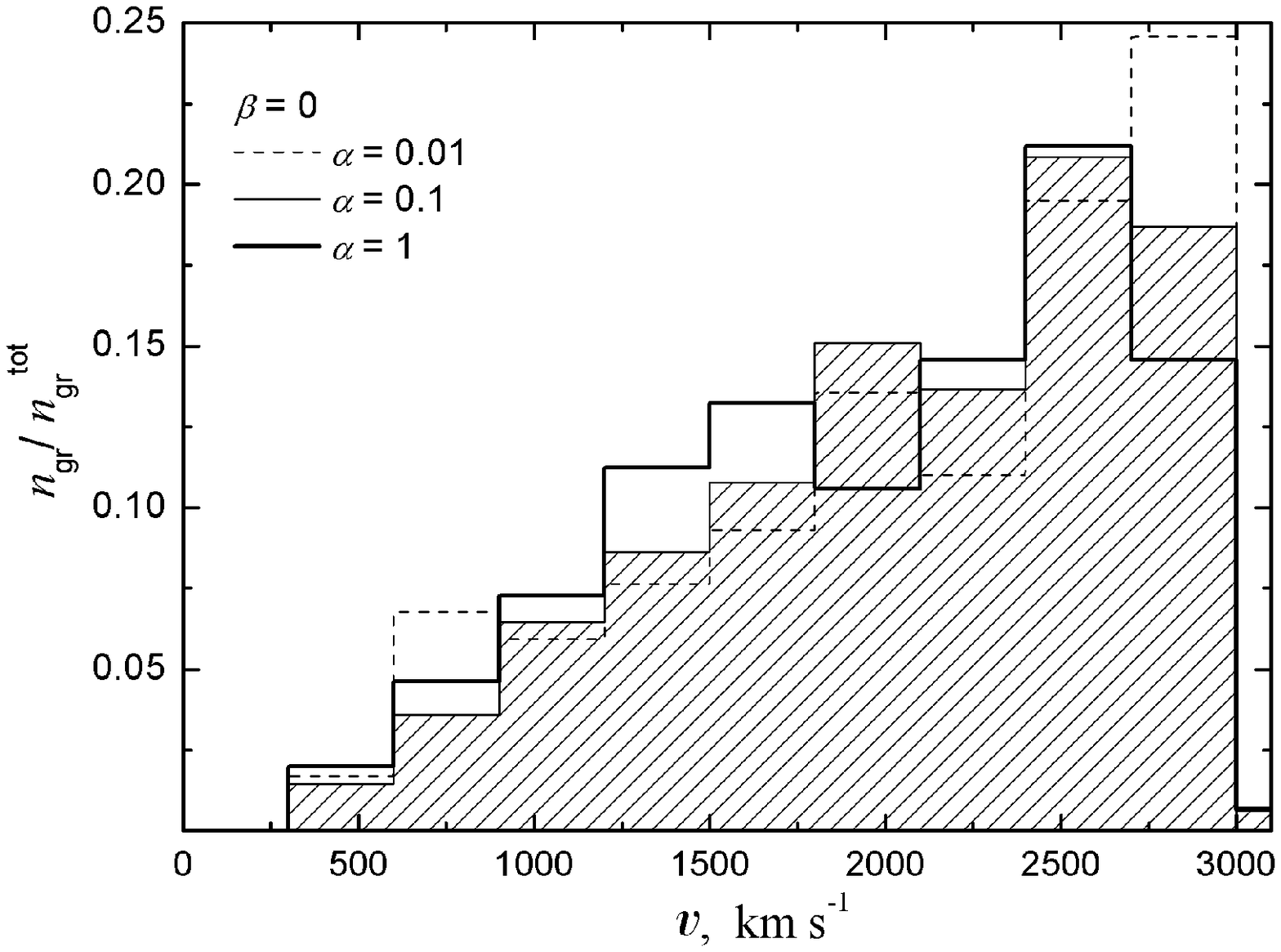} &
\includegraphics[angle=0, width=0.465\textwidth]{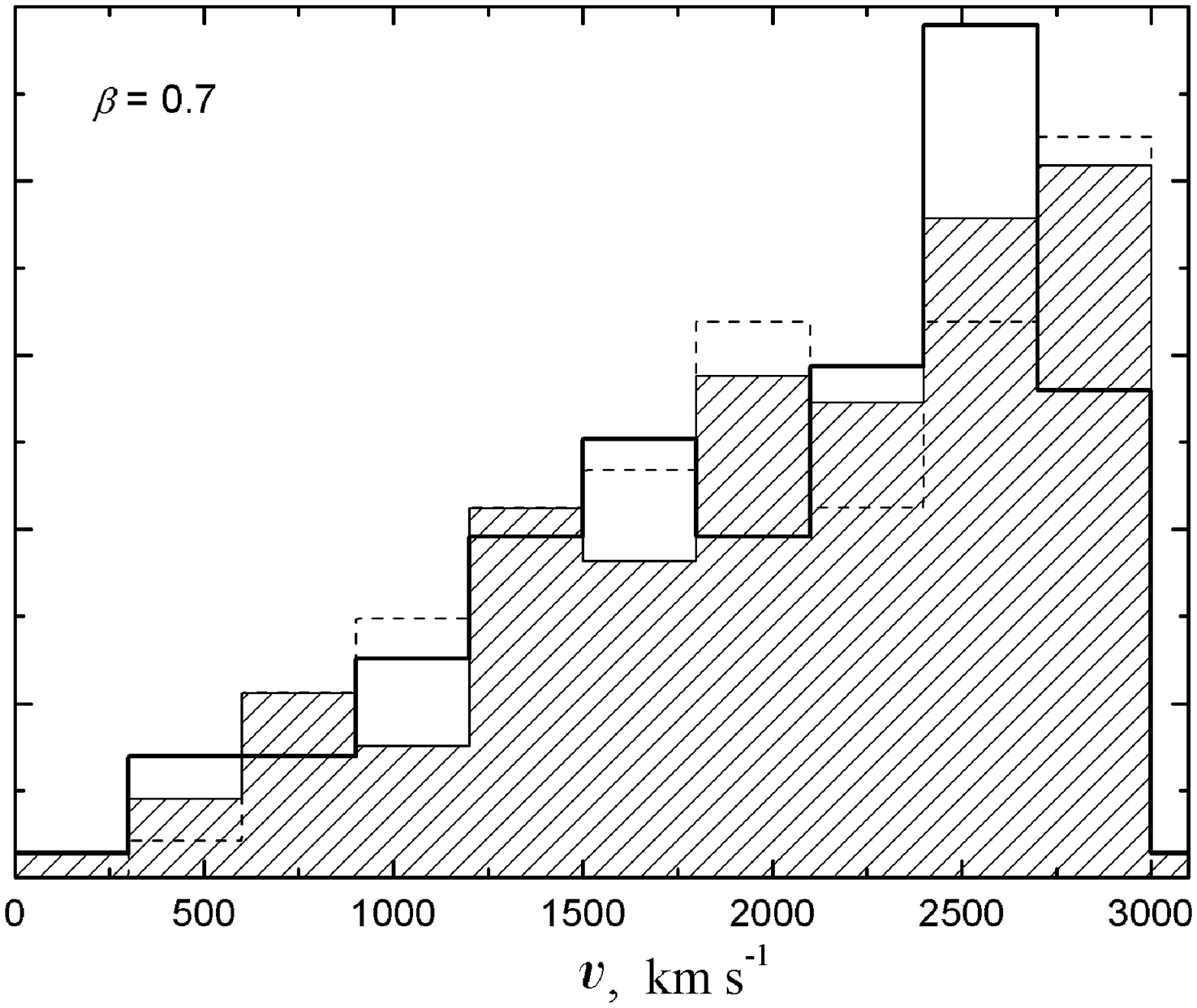} \\

\end{tabular}
\caption{Distributions of galaxies in the MK groups and the MEV subsamples in radial velocity: (a) at ${\beta}$ = 0, (b) at ${\beta}$ = 0.7 and (c and d) the same for the coincided MK and MEV groups.}
\end{figure}

Our group identification shows that it is hard to conclude which of the MEV subsamples coincides most closely with the MK groups, since the number of coincidences for different numbers of galaxies in the group is largest at different values of the parameters ${\alpha}$ and ${\beta}$, as would be expected. The less populated the group, the more such groups coincide. More specifically, groups with a small number of members constitute the largest number of coincided groups: there are about 45 \% of coincided pairs, 30 \% of triplets, and 25 \% of groups with 4 to 10 components.

The fractions of the galaxies in the MK groups with a population from 2 to 10 components and more than 10 components are, respectively, 39 and 14 \% (a total of 53 \% of the clustered ones). 26, 28, and 30 \% of the total number of galaxies at ${\beta}$ = 0 and ${\alpha}$ = 0.01, 0.1, and 1 \%, respectively, fell into the MEV groups with a population from 2 to 10 components; 10, 18, and 28 \% of the total number of galaxies fell into the more populated groups ($>$10 components).

This suggests that the Voronoi tessellation method depends weakly on ${\beta}$ and that the number of galaxies in rich structures increases with ${\alpha}$ much faster than in poor ones. On the whole, the Karachenstev and Voronoi methods tend to identify sparsely and more densely populated groups, respectively. Therefore, the total numbers of MK and MEV groups differ, on average, by 300 (see Table 3).

Figures 9a and 9b present the galaxy distributions for the groups being compared in radial velocity. We see that the galaxy distribution in the MEV subsamples agrees well with that in the MK groups, i.e., there is no supercluster depth selection in this case. At the same time, the number of coincided groups is larger on the supercluster periphery (Figs. 9c and 9d). One of the explanations is that a strong attractor, the Virgo cluster of galaxies, is located at a distance of about 1200 km s$^{-1}$. Therefore, the group identification in this sparsely populated region depends strongly on the clustering method and parameters and requires additional information about the individual characteristics of galaxies. For this reason, the smallest number of MK and MEV groups coincide by all components in the Virgo cluster.

\section{DISCUSSION AND RESULTS}

We used the 3D Voronoi tessellation method to identify groups of galaxies in the Local Supercluster and compared the derived groups with MK groups [15] using the same catalog of LS galaxies (N = 7064).

The 3D Voronoi tessellation method is a geometrical method based only on the positions of galaxies in space. The method reveals regions with an enhanced galaxy density compared to the background (i.e., a den­sity contrast compared to a random distribution). Since the LS catalog is inhomogeneous (there is a strong selection of dwarf galaxies with depth), we artificially rescaled the distances in such a way that the concen­tration of galaxies varied with sample depth as a power law with the same index ${\beta}$ as that for the full homo­geneous catalog. In turn, the Karachentsev method is dynamical and takes into account the individual prop­erties of galaxies. It is based on the assumption about circular Keplerian motions of the group components around the main (most massive) galaxy; the group boundary is its so-called zero-velocity surface [3, 12].

Analysis of the Voronoi cell sizes in the LS showed the following: the more stringently the galaxy catalog is limited in $M_{abs}$, the more information about the distribution of real galaxies is lost. Luminous galaxies trace weakly the distribution of galaxy groups of the full catalog. The distribution of the most luminous gal­axies, $M_{abs} < -20.5^{m}$, is closer to a random one (but does not correspond to it). However, we would like to emphasize that this effect in the distribution of galaxies may depend not on their luminosity, but on their number: the fewer the galaxies, the weaker their clustering (see Figs. 5 and 6).

We identified the MEV groups derived by the 3D Voronoi tessellation method at various clustering parameters (${\alpha}$ and ${\beta}$) with the MK groups found by the dynamical method. The fraction of the MEV and MK groups that coincided by all components is 22 \%, which is quite acceptable. For example, when the hier­archical and percolation methods were compared, 25 \% of the groups coincided [6]. Since the 3D Voronoi tessellation method has been used for the first time to solve problems of this kind, this overall statistical coincidence between the applications of different methods suggests that its use for identifying galaxy groups is legitimate. At the same time, it cannot be concluded what values of the parameters ${\alpha}$ and ${\beta}$ are optimal, since the number of coincidences for different numbers of components in the group is largest at different values of ${\alpha}$ and ${\beta}$.The Voronoi tessellation method depends weakly on ${\beta}$ and the number of gal­axies in rich structures increases with ${\alpha}$ much faster than in poor ones. Comparison of the coincided groups obtained by different clustering methods for the same LS galaxy sample shows that the Karachentsev and Voronoi tessellation method tend to identify sparsely and more densely populated groups, respectively. The advantage of the Voronoi tessellation method is its simplicity, since only the coordinates and velocities of galaxies must be known. Thus, the method can be useful in a preliminary (initial) study of the structure of a supercluster or when there is no information about the individual properties of galaxies. The distance rescaling method is also of great importance, since any other geometrical clustering method can be used after its application. Determining the physical characteristics of galaxy groups is important both for elucidating the LS structure and for estimating the mass-to-light ratio, since it is a unique source of data on the distri­bution of dark matter on scales 0.1-1 Mpc [12]. When the kinematical and dynamical characteristics of gal­axy groups are compared, the quality of the original sample and the criterion for selection in the groups are of crucial importance for choosing a particular dark matter distribution model [1,2]. This suggests that it would be appropriate to use several methods to identify the galaxy supercluster structure, since in combina­tion these give a weighty help in the observational programs to search for and identify members of physi­cally associated groups.

\bigskip

{\bf ACKNOWLEDGMENTS.} We wish to thank D.I. Makarov and I. D. Karachentsev, who provided the LS galaxy catalog before its publication and for helpful discussions. This work was supported by the State Foundation for Basic Research of the Ministry of Education and Science of Ukraine (grant no. F7/267-2001).
We used the LEDA database (http://leda.univ-lyon1.fr) and NED (http://nedwww.ipac.caltech.edu). 

{}
\end{document}